\documentclass[twocolappendix,appendixfloats]{emulateapj}
\usepackage{apjfonts}
\usepackage{rotating}
\usepackage{amsmath}
\usepackage{color}

\newcommand{\pivec}{\mbox{\boldmath $\pi$}}

\lefthead{HAN ET AL.} 
\righthead{OGLE-2014-BLG-0257L}

\begin{document}
\title{OGLE-2014-BLG-0257L: A Microlensing Brown Dwarf Orbiting a Low-mass M Dwarf}

\author{
C.~Han$^{1}$, Y.~K.~Jung$^{1}$, A.~Udalski$^{2,7}$, A.~Gould$^{3,4}$, V.~Bozza$^{5,6}$, \\
and\\
M.~K.~Szyma{\'n}ski$^{2}$, 
I.~Soszy{\'n}ski$^{2}$, 
R.~Poleski$^{2,3}$,
S.~Koz{\l}owski$^{2}$, 
P.~Pietrukowicz$^{2}$, 
J.~Skowron$^{2}$,
K.~Ulaczyk$^{2}$, 
{\L}.~Wyrzykowski$^{2}$\\
(The OGLE Collaboration),\\
}

\affil{$^{1}$ Department of Physics, Institute for Astrophysics, Chungbuk National University, 371-763 Cheongju, Korea}
\affil{$^{2}$ Warsaw University Observatory, Al. Ujazdowskie 4, 00-478 Warszawa, Poland}
\affil{$^{3}$ Department of Astronomy, Ohio State University, 140 W. 18th Ave., Columbus, OH  43210, USA} 
\affil{$^{4}$ Max-Planck-Institute for Astronomy, K\"onigstuhl 17, 69117 Heidelberg, Germany}
\affil{$^{5}$ Dipartimento di Fisica "E.R. Caianiello" Universit$\grave{\rm a}$ degli Studi di Salerno Via Giovanni Paolo II - I 84084 Fisciano (SA), Italy}
\affil{$^{6}$ Istituto Nazionale di Fisica Nucleare, Sezione di Napoli, Italy} 
\footnotetext[7]{The OGLE Collaboration}

\begin{abstract}
In this paper, we report the discovery of a binary composed of a brown dwarf and a low-mass 
M dwarf from the observation of the microlensing event OGLE-2014-BLG-0257. Resolution of the 
very short-lasting caustic crossing combined with the detection of subtle continuous deviation 
in the lensing light curve induced by the Earth's orbital motion enable us to precisely 
measure both the Einstein radius $\theta_{\rm E}$ and the lens parallax $\pi_{\rm E}$, 
which are the two quantities needed to unambiguously determine the mass and distance to the 
lens. It is found that the companion is a substellar brown dwarf with a mass $0.036\pm 0.005\ M_\odot$ 
($37.7\pm 5.2\ M_{\rm J}$) and it is orbiting an M dwarf with a mass $0.19\pm 0.02\ M_\odot$. 
The binary is located at a distance $1.25\pm 0.13$ kpc toward the Galactic bulge and the projected 
separation between the binary components is $0.61\pm 0.07$ AU.  The separation scaled by the mass 
of the host is $3.2\ {\rm AU}/M_\odot$.  Under the assumption that separations scale with masses, 
then, the discovered brown dwarf is located in the zone of the brown dwarf desert.  With the increasing 
sample of brown dwarfs existing in various environments, microlensing will provide a powerful probe 
of brown dwarfs in the Galaxy.
\end{abstract}


\section{Introduction}

Since the first discoveries of Teide 1 \citep{rebolo1995} and Gliese 229B \citep{nakajima1995} in 1995, 
the last two decades have witnessed a wealth of brown dwarf (BD) discoveries. 
In the archives ``{http://DwarfArchives.org}'' and the one managed by J.~Gagne\footnote{ 
https://jgagneastro.wordpress.com/list-of-ultracool-dwarfs}, there are thousands of known objects 
fall in the temperature regime occupied by brown dwarfs.

\begin{deluxetable*}{lllll}
\tablecaption{Microlensing brown dwarfs\label{table:one}}
\tablewidth{0pt}
\tablehead{
\multicolumn{1}{c}{Lensing event} &
\multicolumn{1}{c}{Lens}          &
\multicolumn{2}{c}{Mass}          &
\multicolumn{1}{c}{Reference}     \\
\multicolumn{1}{c}{}              &
\multicolumn{1}{c}{}              &
\multicolumn{1}{c}{Primary}       &
\multicolumn{1}{c}{Secondary}     &
\multicolumn{1}{c}{} 
}
\startdata
OGLE-2006-BLG-277                              & BD around an M dwarf         & $0.10\pm 0.03\ M_{\odot}$   & $52\pm 15\ M_{\rm J}$     & (1)   \\ 
MOA-2007-BLG-197                               & BD around a G-K dwarf        & $0.82\pm 0.04\ M_{\odot}$   & $41\pm 2\ M_{\rm J}$      & (2)   \\
OGLE-2007-BLG-224                              & Isolated BD                  & $59\pm 4\ M_{\rm J}$        & -                         & (3)   \\
OGLE-2008-BLG-510\footnote{MOA-2008-BLG-369}   & BD around an M dwarf         & -                           & -                         & (4)   \\
OGLE-2009-BLG-151\footnote{MOA-2009-BLG-232}   & Binary system of BDs         & $19.1\pm 1\ M_{\rm J}$      & $7.9\pm 0.3\ M_{\rm J}$   & (5,6) \\
MOA 2009-BLG-411                               & BD around an M dwarf         & $0.18\pm 0.02\ M_{\odot}$   & $53\pm 5\ M_{\rm J}$      & (7)   \\  
MOA-2010-BLG-073                               & BD around an M dwarf         & $0.16\pm 0.03\ M_{\odot}$   & $11.0\pm 2.0\ M_{\rm J}$  & (8)   \\
OGLE-2011-BLG-0420                             & Binary system of BDs         & $26.1\pm 1\ M_{\rm J}$      & $9.9\pm 0.5\ M_{\rm J}$   & (6)   \\
MOA-2011-BLG-104\footnote{OGLE-2011-BLG-0172}  & BD around an M dwarf         & $0.18\pm 0.11\ M_{\odot}$   & $21\pm 10\ M_{\rm J}$     & (9)   \\
MOA-2011-BLG-149                               & BD around an M dwarf         & $0.14\pm 0.02\ M_{\odot}$   & $20\pm 2\ M_{\rm J}$      & (9)   \\
OGLE-2012-BLG-0358                             & BD hosting a planet          & $23.2\pm 2.0\ M_{\rm J}$    & $1.9\pm 0.2\ M_{\rm J}$   & (10)  \\
OGLE-2013-BLG-0102                             & BD around an M dwarf         & $0.10\pm 0.01\ M_{\odot}$   & $12.6\pm 2.1\ M_{\rm J}$  & (11)  \\ 
OGLE-2013-BLG-0578                             & BD around an M dwarf         & $0.124\pm 0.014\ M_{\odot}$ & $33.6\pm 4.2\ M_{\rm J}$  & (12)  \\
OGLE-2015-BLG-1268                             & Isolated BD                  & $47\pm 7\ M_{\rm J}$        & -                         & (13)   
\enddata                                                                                              
\tablecomments{
(1) \citet{park2013}, (2) \citet{ranc2015}, (3) \citet{gould2009}, (4) \citet{bozza2012}, 
(5) \citet{shin2012a}, (6) \citet{choi2013}, (7) \citet{bachelet2012}, (8) \citet{street2013}, 
(9) \citet{shin2012c}, (10) \citet{han2013}, (11) \citet{jung2015}, (12) \citet{park2015}, 
(13) \citet{zhu2015}.
}                                                      
\end{deluxetable*}                                                                                    

Despite the large number of known BDs, our understanding about their formation processes is still
not clear. As a result, there exist various theories including turbulent fragmentation of molecular 
clouds \citep{boyd2005}, fragmentation of unstable accretion disks \citep{stamatellos2007}, 
ejection of protostars from prestellar cores \citep{reipurth2001}, photo-erosion of prestellar 
cores by nearby very bright stars \citep{whitworth2004}, etc. It may be that these theories apply 
to different populations of BDs formed under different environments. For comprehensive studies of 
the formation mechanism, therefore, it is important to have BD samples detected by using various 
methods that are sensitive to different populations of BDs.

Due to the nature of detecting objects through their gravitational field rather than their radiation,  
microlensing experiments can enrich BD samples by detecting BDs that are difficult to be detected by 
other methods such as old and thus faint or dark brown dwarfs. There have been discovery reports of 
various types of microlensing BDs as listed in Table 1. These samples include an isolated BD 
\citep{gould2009}, BD companions to faint stars \citep{bozza2012, shin2012a, shin2012c, bachelet2012, 
street2013, park2013, jung2015, park2015, ranc2015}, BDs hosting planets \citep{han2013}, and a 
binary system of BDs \citep{choi2013}.

Although rapidly increasing, the number of known microlensing BDs is still limited due to the 
difficulties in detection. The most important obstacle comes from the difficulties 
in identifying the BD nature of the lens by measuring its mass. For general lensing events, the 
only observable parameter related to the mass of the lens is the event time scale $t_{\rm E}$. However, 
the time scale results from the combination of the lens mass, distance, and relative lens-source 
speed, and thus the mass estimated from $t_{\rm E}$ is highly degenerate. 

For the unique 
measurement of the mass, one needs to measure two additional quantities of the angular Einstein 
radius $\theta_{\rm E}$ and the lens parallax $\pi_{\rm E}$. The Einstein radius is measured by 
detecting the effect of the finite source size on the lensing light curve. On the other hand, 
the lens parallax can be measured by detecting long-term deviations in lensing light curves 
caused by the deviation of the lens-source relative motion from rectilinear due to the change 
of the observer's position induced by the orbital motion of the Earth around the Sun 
\citep{gould1992}. With the measured $\theta_{\rm E}$ and $\pi_{\rm E}$, the lens mass and 
distance are determined by
\begin{equation}
M={\theta_{\rm E} \over \kappa \pi_{\rm E}}\qquad
D_{\rm L}={ {\rm AU}\over \pi_{\rm E}\theta_{\rm E}+\pi_{\rm S}},
\end{equation}
respectively. Here $\kappa=4G/(c^2{\rm AU})$, $\pi_{\rm S}={\rm AU}/D_{\rm S}$ is the parallax 
of the source star located at a distance $D_{\rm S}$ \citep{gould2000}.

\begin{figure*}[ht]
\epsscale{0.80}
\plotone{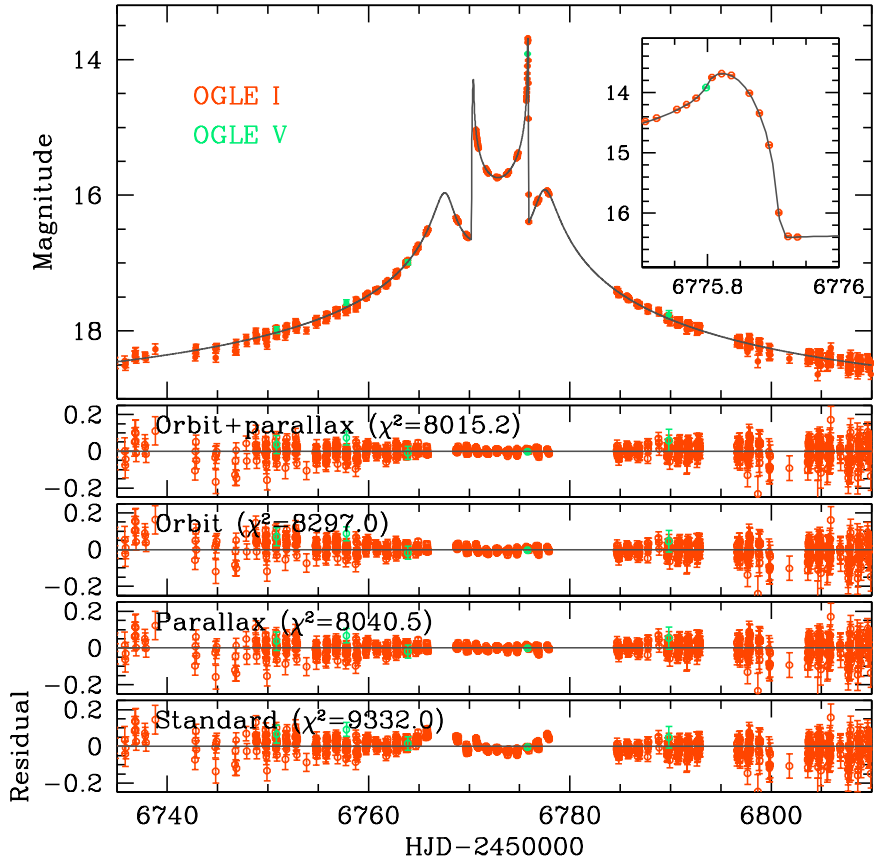}
\caption{\label{fig:one}
Light curve of OGLE-2014-BLG-0257. The inset in the top panel shows the enlarged view around 
the caustic exit. The 4 lower panels show the residuals from the 4 tested models.}
\end{figure*}

For a lensing event produced by a single mass, finite-source effects can be detected only for 
rare cases of high-magnification events where the lens approaches the source star close 
enough to pass over the source star surface \citep{witt1994, gould1994, nemiroff1994, choi2012}.
The chance to measure the lens parallax is also very 
slim because events produced by low-mass BDs tend to have short time scales while the Earth's 
parallactic motion has long-term effects on lensing light curves.

In contrast to single-lens events, the chance to measure lens masses is much higher for events 
produced by binary lenses. This is first because the majority of binary-lens events are involved with 
caustic crossings during which finite-source effects are important and thus the chance to 
measure $\theta_{\rm E}$ is high  \citep{dominik1995, gaudi1999, gaudi2002, pejcha2009}.
Caustics represent the positions on the source plane at which a point-source 
magnification is infinite and they form a single or multiple closed curves of extended size.
For a finite-source, the lensing magnification corresponds to the mean averaged over the source star surface
and thus deviates from that of a point source.
In addition, although masses of BDs are small, binaries 
including BDs can be massive and thus events can have time scales long enough for one to 
measure  $\pi_{\rm E}$. As a result, most known microlensing BDs were detected through the 
channel of binary-lens events, e.g. \citep{shin2012b}.

Another difficulty of microlensing BD detection comes from the fact that BD signals in the lensing 
light curve do not have distinctive features. For a binary lens with an extreme companion/primary 
mass ratio $q$ like a planet-star pair, the caustic induced by the companion is very tiny and 
the signal of the companion appears as a distinctive short-term anomaly that is superposed on 
the smooth and symmetric lensing light curve of the primary \citep{gaudi2012}.
For BD companions with mass ratios 
$q\sim 0.1$, the size of the induced caustic can be as large as those induced by roughly 
equal-mass binaries.  As a result, it is difficult to immediately notice that the mass of the 
companion is in the regime of BDs and thus identifying BD companions requires complex 
procedure of modeling observed light curves.

In this paper, we report the discovery of a BD that is bound to a low-mass star. The BD nature 
of the companion is identified from the mass measurement, which is possible due to the resolution 
of a very short-term caustic crossing combined with the detection of a subtle long-term 
deviation in the lensing light curve induced by the Earth's orbital motion.

\section{Observation and data}

The binary system was detected from the observation of a microlensing event OGLE-2014-BLG-0257. 
The event occurred on a star that lies toward the Galactic Bulge field with equatorial and Galactic 
coordinates $(\alpha,\delta)_{\rm J2000}=(18^{\rm h}01^{\rm m}47.79^{\rm s}, -28^\circ 15'43.2'')$ 
and $(l,b)_{\rm J2000}=(2.38^\circ, -2.72^\circ)$, respectively. The lensing-induced brightening of 
the star was first noticed in April, 2014 by the Optical Gravitational Lensing Experiment 
\citep[OGLE-IV:][]{udalski2015}
survey that is conducted using the 1.3-m Warsaw telescope located at the Las Campanas Observatory 
in Chile. 
In Figure~\ref{fig:one}, we present the event light 
curve which shows a typical binary-lensing feature of caustic-crossing spikes with a ``U``-shaped trough 
between them. There also exist two bumps that occurred before and after the caustic crossings.

With the progress of the event, it was noticed that the magnification of the lensed star (source) 
flux was high. Since high-magnification events are sensitive to planetary companions to lenses 
\citep{griest1998}, a second alert was issued to the microlensing community on April, 18 
(${\rm HJD}\sim 2456765$) for intensive follow-up observation. On April 23 (${\rm HJD}\sim 2456770$), 
a strong anomaly from a single lensing light curve was detected. Such an anomaly is usually associated 
with a caustic formed by a binary lens. Caustics form closed curves and thus caustic 
crossings of a source star occur in pairs, i.e. caustic entrance and exit. Since resolving caustic 
crossings is important to detect finite-source effects and thus to measure the Einstein radius of the 
lens, another alert was issued. Despite the successive alerts, immediate follow-up observation could 
not be done since the anomaly occurred during the early Bulge season when the time window of the 
bulge visibility was narrow and thus telescopes for follow-up observations were not fully operational. 
Fortunately, the source star was located in the field toward which stars were monitored 
with a very high cadence ($\gtrsim 3\ {\rm hr}^{-1}$) and thus the caustic exit was densely resolved 
despite the short duration ($\sim 3.5$ hours) of the caustic crossing. See the inset 
in the upper panel of Figure~\ref{fig:one}, for the enlarged caustic-crossing part of the light curve.

After the caustic exit, there was another bump induced by the source star's approach close to a 
cusp of the caustic. From the analyses of the light curve conducted after the second bump, it was 
noticed that the mass ratio between the binary lens components is $q\sim 0.1$ -- 0.2, indicating 
that the mass of the companion is very low.  From continued analyses conducted with the progress of 
the event, it was also noticed that long-term higher-order effects were needed to precisely describe 
the observed light curve. The event lasted for about $\sim 3$ months before it returned to its baseline 
magnitude $I\sim 19$.

In the analysis, we use 8031 OGLE $I$-band data points and
107 $V$-band data points.  Photometry of data were conducted by using the 
customized pipeline \citep{udalski2003} that is based on the Difference Imaging Analysis code 
developed by \citet{wozniak2000}.
For the use of analysis, we readjust the flux measurement uncertainties of each 
data set by first adding a quadratic term to make the cumulative $\chi^2$ distribution ordered by 
magnifications approximately linear and then rescaling the uncertainties so that $\chi^2$ per degree 
of freedom (dof) becomes unity for the best-fit model.

\begin{deluxetable*}{lrrrrrr}
\tablecaption{Lensing parameter\label{table:two}}
\tablewidth{0pt}
\tablehead{
\multicolumn{1}{c}{Parameter} & 
\multicolumn{1}{c}{Standard}  & 
\multicolumn{2}{c}{Parallax}  &
\multicolumn{1}{c}{Orbital}  & 
\multicolumn{2}{c}{Orbital + Parallax} \\
\multicolumn{1}{c}{} &
\multicolumn{1}{c}{} &
\multicolumn{1}{c}{$u_0>0$} &
\multicolumn{1}{c}{$u_0<0$} &
\multicolumn{1}{c}{} &
\multicolumn{1}{c}{$u_0>0$} &
\multicolumn{1}{c}{$u_0<0$} 
}
\startdata
$\chi^2$                       & 9332.0               &  8040.5               &   8044.1              & 8297.0               &   8015.2               &    8025.4              \\
$t_0$ (HJD')                   & $6774.233\pm 0.021$  &  $6774.274\pm 0.017$  &  $6774.255\pm 0.017$  & $6774.206\pm 0.018$  &   $6774.314\pm 0.019$  &  $ 6774.321\pm 0.019$  \\   
$u_0$ ($10^{-2}$)              & $0.83    \pm 0.01 $  &  $0.78    \pm 0.01 $  &  $-0.82   \pm 0.01 $  & $0.92    \pm 0.01$   &   $0.81    \pm 0.02 $  &  $-0.80    \pm 0.01 $  \\
$t_{\rm E}$ (days)             & $68.30   \pm 0.11 $  &  $77.81   \pm 0.23 $  &  $74.80   \pm 0.56 $  & $66.05   \pm 0.36$   &   $77.93   \pm 1.40 $  &  $ 78.73   \pm 1.11 $  \\
$s$                            & $0.44    \pm 0.01 $  &  $0.42    \pm 0.01 $  &  $0.43    \pm 0.01 $  & $0.43    \pm 0.01$   &   $0.43    \pm 0.01 $  &  $ 0.43    \pm 0.01 $  \\
$q$                            & $0.22    \pm 0.01 $  &  $0.20    \pm 0.01 $  &  $0.21    \pm 0.01 $  & $0.25    \pm 0.01$   &   $0.19    \pm 0.01 $  &  $ 0.18    \pm 0.01 $  \\
$\alpha$ (rad)                 & $3.116   \pm 0.001$  &  $3.132   \pm 0.001$  &  $-3.133  \pm 0.001$  & $3.130   \pm 0.001$  &   $3.132   \pm 0.001$  &  $-3.132   \pm 0.001$  \\
$\rho$ ($10^{-3}$)             & $0.61    \pm 0.01 $  &  $0.52    \pm 0.01 $  &  $0.54    \pm 0.01 $  & $0.61    \pm 0.01$   &   $0.53    \pm 0.01 $  &  $ 0.520   \pm 0.01 $  \\
$\pi_{{\rm E},N}$              & -                    &  $0.71    \pm 0.02 $  &  $-0.73   \pm 0.02 $  & -   -                &   $0.59    \pm 0.06 $  &  $-0.76    \pm 0.06 $  \\
$\pi_{{\rm E},E}$              & -                    &  $-0.11   \pm 0.01 $  &  $-0.20   \pm 0.01 $  & -   -                &   $-0.13   \pm 0.01 $  &  $-0.22    \pm 0.01 $  \\
$ds/dt$ (${\rm yr}^{-1}$)      & -                    &    -                  &  -                    & $-0.38   \pm 0.04$   &   $0.19    \pm 0.05 $  &  $ 0.25    \pm 0.05 $  \\
$d\alpha/dt$ (${\rm yr}^{-1}$) & -                    &    -                  &  -                    & $1.99    \pm 0.07$   &   $0.78    \pm 0.25 $  &  $-0.09    \pm 0.22 $ 
\enddata                                                                                            
\tablecomments{${\rm HJD}'={\rm HJD}-2450000$.}
\end{deluxetable*}

\section{Modeling}

The basic description of a binary-lensing light curve requires seven lensing parameters: $t_0$, 
$u_0$, $\alpha$, $t_{\rm E}$, $s$, $q$, and $\rho$. The first four of these parameters describe 
the source trajectory with respect to the lens: $t_0$ is the time of the source star's closest 
approach to a reference position in the binary lens, $u_0$ is the separation between the source 
and the reference position at $t_0$, $\alpha$ is the angle between the source trajectory and the 
binary axis, and $t_{\rm E}$ is the time scale required for the source to cross the Einstein radius 
corresponding to the total mass of the binary. For a reference position on the lens plane, we use 
the center of mass of the binary lens. Two other parameters describe the binary lens: $s$ is the 
projected separation normalized to $\theta_{\rm E}$, and $q$ is the mass ratio of the lens components. 
The last basic parameter, $\rho$, is the source radius normalized to $\theta_{\rm E}$. The normalized 
source radius is needed to describe the caustic-crossing part of the light curve that is affected 
by finite-source effects.

The basic features of binary lensing light curves are determined by the size and shape of the 
caustic, which depends on $s$ and $q$, and the source trajectory with respect to the caustic, 
which depends on $\alpha$. In the initial modeling run, we therefore explore the $(s, q, \alpha)$ 
parameter space by conducting a grid search, while the other parameters are searched for by 
minimizing $\chi^2$.  For the $\chi^2$ minimization, we use the Markov Chain Monte Carlo (MCMC) 
method. From a $\chi^2$ map in the $(s, q, \alpha)$ space obtained from this initial run, we 
identify the approximate locations of the local minima. We then perform a $\chi^2$ optimization 
using all parameters at each local minima and refine the solution. Finally, we identify a global 
minimum by comparing $\chi^2$ values of the individual local minima. The grid search is important 
to identify the existence of degenerate solutions which result in similar light curves despite the 
combinations of dramatically different parameters.

For the computation of finite-source lensing magnifications around the caustic crossings, we 
use the numerical ray-shooting method. In this process, we consider limb-darkening effects 
of the source star by modeling its surface brightness profiles as $S_\lambda \propto 1-\Gamma_\lambda 
(1-3\cos \phi/2)$, where $\Gamma_\lambda$ is the linear limb-darkening coefficient and $\phi$ is 
the angle between the light of sight toward the center of the source and the normal to the source 
surface \citep{albrow1999}.
From the de-reddened color and brightness, we find that the source is an F-type main 
sequence (see section 4 for more details) and adopt a limb-darkening coefficient $\Gamma_I=0.37$
from the catalog of \citet{claret2000}. 
In the grid search, we use the map-making method \citep{dong2006}, where a magnification map 
for a given $(s,q)$ is constructed to produce many light curves resulting from different source 
trajectories.

We identify two local minima in the close ($s<1$) and wide ($s>1$) binary regimes resulting from 
the well known close/wide binary degeneracy \citep{dominik1999, bozza2000, an2005}. However, the 
degeneracy is not severe and the close binary solution provides a better fit with a significant 
confidence level ($\Delta\chi^2 \gtrsim 400$). From the modeling run based on the basic parameters 
(standard model), we find that the event was produced by a close binary with a projected 
separation $s\sim 0.4$ and a mass ratio $q\sim 0.2$.

Although the standard model explains the main features of the light curve, it is found that 
there exist noticeable residuals lasting for $\sim 20$ days from the first bump through the 
caustic-crossing feature to the second bump. See the residual from the standard model at the 
bottom panel of Figure~\ref{fig:one}. These residuals may indicate the need to consider 
higher-order effects. It is known that such a continuous deviation can be caused by either the 
orbital motion of the Earth around the Sun \citep[``parallax effect'':][]{refsdal1966, gould1992}
and/or the change of the lens position caused by the orbital motion of the lens \citep[``lens 
orbital effect'':][]{albrow2000, dominik1998}.\footnote{Besides the parallax and lens-orbital
effects, long-term deviations can also be produced by the orbital motion of the source star if 
it is a binary. We discuss this effect in the Appendix.} 
In order to check these higher-order effects, we 
test additional models. In the ``parallax'' and ``orbital'' models, we consider the parallax and 
lens-orbital effects, respectively. In the ``orbital+parallax'' model, we consider both higher-order 
effects. Consideration of the parallax effect requires to include two additional parameters 
$\pi_{{\rm E},N}$ and $\pi_{{\rm E},E}$, that are the two components of the lens parallax vector 
$\pivec_{\rm E}$ projected onto the sky along the north and east equatorial coordinates, respectively. 
To first-order approximation, the lens-orbital effects are described by two parameters $ds/dt$ and 
$d\alpha/dt$, that are the change rates of the projected binary separation and the source trajectory 
angle, respectively.
We measure the parallax parameters with respect to the reference time that coincides with $t_0$.

\begin{figure}[tbh]
\epsscale{1.1}
\plotone{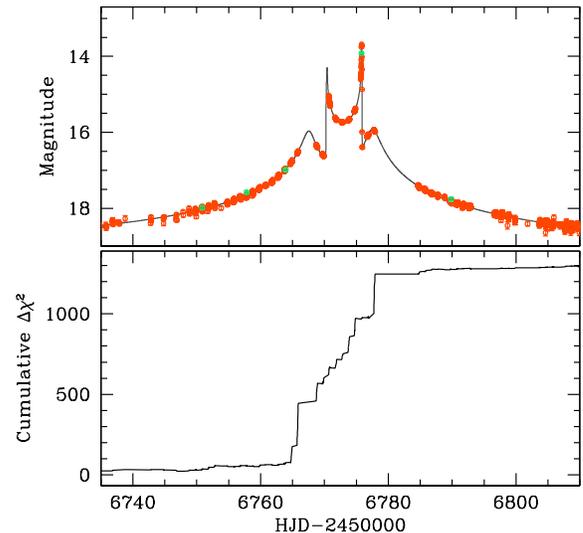}
\caption{\label{fig:two}
Cumulative distribution of $\chi^2$ difference between the best-fit 
(orbital+parallax) and the standard binary models. 
}
\end{figure}

It is known that a pair of parallax solutions with $u_0>0$ and $u_0<0$ result in similar 
light curves due to the mirror symmetry of the source trajectory with respect to the 
binary axis \citep{skowron2011}. 
This so-call ``ecliptic degeneracy'' is important especially for events that occur on 
source stars located near the ecliptic plane. The source star of OGLE-2014-BLG-0257 is 
located just $4.8^\circ$ away from the ecliptic plane and thus this degeneracy might be 
important. We therefore consider both $u_0>0$ and $u_0<0$ solutions whenever we consider 
parallax effects. We note that the pair of the two solutions resulting from the ecliptic 
degeneracy have similar lensing parameters except for the opposite signs of $u_0$, $\alpha$, 
$\pi_{{\rm E},N}$, and $d\alpha/dt$.

\begin{figure}[t]
\epsscale{1.1}
\plotone{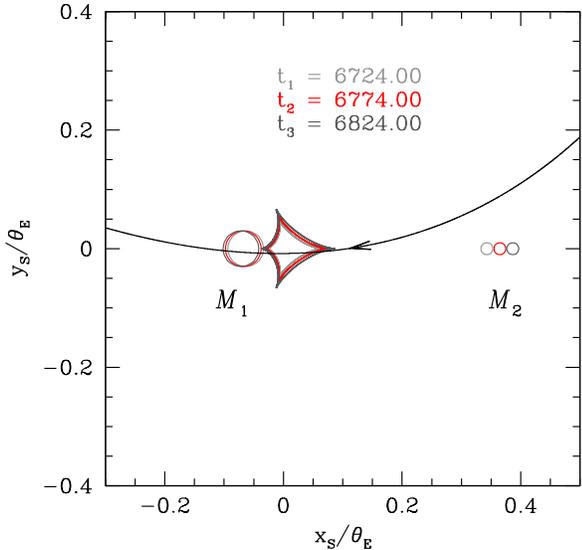}
\caption{\label{fig:three}
Source trajectory (curve with an arrow) with respect to the caustic (cuspy closed figure). 
The two empty circles marked by $M_1$ and $M_2$ represent the positions of the binary 
lens components, where $M_1 > M2$. We note that the positions of the lens components and 
the caustic vary in time due to the orbital motion of the binary lens and thus we mark 
locations of the lens components and the caustic at three different times that are marked 
in the panel. The abscissa is parallel with the binary axis and all lengths are normalized 
to the angular Einstein radius corresponding to the total mass of the binary, $\theta_{\rm E}$.
}
\end{figure}

In Table~\ref{table:two}, we present solutions of the lensing parameters for the tested 
models along with the $\chi^2$ values of the fits. The uncertainties of the individual 
parameters are estimated based on the distributions of points on the MCMC chain. It is 
found that the higher-order effects significantly improve the fit. When the parallax and 
orbital effects are separately considered, the fit improves by $\Delta\chi^2=1291.5$ and 
1035.0, respectively, compared to the standard model. When both effects are simultaneously 
considered, the improvement is $\Delta\chi^2=1316.8$.  See the decrease of the residual 
for the individual models presented in the lower panels of Figure~\ref{fig:one}. 
Considering that (1) the improvement by the parallax effect is significantly greater 
than the orbital effect and (2) the $\chi^2$ difference between the parallax-only and 
parallax+orbital models ($\Delta\chi^2=25.3$) is minor, we judge that the parallax 
effect plays a dominant role in the fit improvement. In Figure~\ref{fig:two}, we present 
the cumulative distribution of $\Delta\chi^2$ of the best-fit model (orbit+parallax model) 
with respect to the standard model. It shows that the improvement occur during the major
anomalies of the bumps and caustic-crossing features. We find that the two solutions caused 
by the ecliptic degeneracy is quite severe although the $u_0>0$ solution is preferred over 
the $u_0<0$ solution by $\Delta\chi^2=10.2$. Since this level of $\Delta\chi^2$ can often 
be ascribed to systematics in data, one cannot completely rule out the $u_0<0$ solution.  
However, we note that the lensing parameters of the two solutions are similar 
except for the signs, and thus the estimated physical parameters are also similar to 
each other.

Figure~\ref{fig:three} shows the geometry of the lens event, where we mark the locations 
of the binary-lens components ($M_1$ and $M_2$), the caustic, and the source trajectory 
with respect to the caustic. We note that the positions of the lens components and the 
caustic vary in time due to the orbital motion of the binary lens and thus we present
locations at three different times that are marked in the panel. It is found that a 
four-cusp central caustic of a close binary lens is responsible for the anomalous features 
in the lensing light curve.  The spikes were produced by the passage of the source 
trajectory through the caustic and the two bumps were produced by the approach of the 
source close to the cusps of the caustic before and after the caustic crossing. Considering 
that only a curved source trajectory can approach both cusps close enough to produce the 
two strong bumps, the clear detection of the parallax effect was possible because of the 
good coverage of the major anomalies. The usefulness of triple-peak (one caustic crossing 
plus two cusp approaches) events in measuring the lens parallax was pointed out by 
\citet{an2001}.

\begin{figure}[t]
\epsscale{1.1}
\plotone{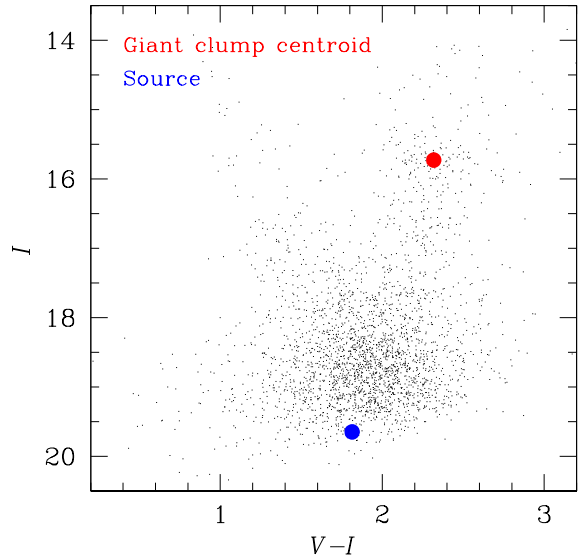}
\caption{\label{fig:four}
Locations of the lensed star and the centroid of giant clump in the 
instrumental color-magnitude diagram.
}
\end{figure}

\section{Physical Quantities}

Among the two quantities needed for the lens mass measurement, the lens parallax 
is measured from modeling. The Einstein radius is determined from the combination 
of the normalized source radius measured from modeling and the angular source 
radius $\theta_*$ by $\theta_{\rm E}=\theta_*/\rho$.

We estimate the $\theta_*$ from the de-reddened color $(V-I)_0$ and brightness 
$I_0$ of the source star. To measure the color and brightness, we first locate 
the source star in the instrumental color-magnitude diagram of neighboring stars 
in the same field and then calibrate the color and brightness by using the centroid 
of the giant clump (GC) as a reference \citep{yoo2004}. The GC centroid can be 
used as a standard candle because its de-reddened color and brightness, 
$(V-I,I)_{0,{\rm GC}}=(1.06,14.35)$ \citep{bensby2011, nataf2013}, are known. 
Figure~\ref{fig:four} shows the locations of the source and GC centroid in the 
instrumental color-magnitude diagram. The measured offsets in color and magnitude 
between the source and the GC are $\Delta(V-I)=-0.50$, $\Delta I=3.92$, respectively. 
Then, the estimated de-reddened color and the brightness of the source star are 
$(V-I,I)_0=(0.56,18.26)$, respectively, indicating that the source is a mid F-type 
main sequence. We then convert the estimated $(V-I)_0$ into $(V-K)_0=1.16$ using 
the color-color relation of \citet{bessell1988} and finally determine $\theta_*$ 
using the color-angular radius relation of \citet{kervella2004}.  We find that the 
angular radius of the source star is $\theta_*=0.59\pm 0.04\ \mu{\rm as}$. Then the 
angular Einstein radius corresponding to the best-fit model (orbital+parallax model 
with $u_0>0$) is 
\begin{equation}
\theta_{\rm E} = {\theta_*\over \rho} = 1.13 \pm 0.08\ {\rm mas}.
\end{equation}
Combined with the Einstein time scale estimated from lens modeling, the relative 
lens-source proper motion is determined as
\begin{equation}
\mu={\theta_{\rm E}\over t_{\rm E}}=5.28\pm 0.38\ {\rm mas}\ {\rm yr}^{-1}.
\end{equation}
We note that the values of $\theta_{\rm E}$ and $\mu$ are very similar for the 
$u_0<0$ solution due to the similarity in the measured parameters of $\rho$ and 
$t_{\rm E}$.

\begin{deluxetable}{lrr}
\tablecaption{Physical parameters\label{table:three}}
\tablewidth{0pt}
\tablehead{
\multicolumn{1}{c}{Quantity} & 
\multicolumn{1}{c}{$u_0>0$}  & 
\multicolumn{1}{c}{$u_0<0$} 
}
\startdata
Primary mass         & $0.19 \pm 0.02\ M_\odot$    & $0.15 \pm 0.02\ M_\odot$    \\      
Companion mass       & $0.036\pm 0.005\ M_\odot$   & $0.027\pm 0.003\ M_\odot$   \\        
                     & $(37.7\pm 5.2 \ M_{\rm J})$ & $(28.3\pm 3.1\ M_{\rm J})$  \\         
Distance to the lens & $1.25 \pm 0.13\ {\rm kpc}$  & $0.98 \pm 0.09\ {\rm kpc}$  \\         
Projected separation & $0.61 \pm 0.07\ {\rm AU}$   & $0.48 \pm 0.04\ {\rm AU}$   \\
KE/PE                & 0.01                        & 0.01           
\enddata
\end{deluxetable}

In Table~\ref{table:two}, we present the physical parameters of the lens estimated from 
the measured $\theta_{\rm E}$ and $\pi_{\rm E}$, including the masses of the individual 
lens components, $M_1$ and $M_2$, the distance to the lens, $D_{\rm L}$, and the projected 
separations between the components, $r_\perp$. Also presented is the ratio of transverse 
kinetic to potential energy of the binary system estimated by, 
\begin{equation}
\left({\rm KE}\over{\rm PE} \right)_\perp = {(r_\perp/{\rm AU})^3 \over 8\pi^2(M/M_\odot)}
\left[ \left( {1\over s} {ds/dt\over {\rm yr}^{-1}} \right)^2 + \left( {d\alpha/dt\over 
{\rm yr}^{-1}} \right)^2 \right],
\end{equation}
where $M=M_1+M_2$ is the total mass of the binary. 
For a bound system, the ratio $({\rm KE}/{\rm PE})_\perp \leq ({\rm KE}/{\rm PE}) < 1$. 
The measured ratio is very small and thus satisfies the bound-system check.

From the estimated physical parameters, we find that the lens is a binary composed of 
a substellar BD with a mass $M_2=0.036\pm 0.005\ M_\odot$ ($37.7 \pm 5.2\ M_{\rm J}$) 
and a low-mass M dwarf with a mass $M_1=0.19\pm 0.02\ M_\odot$.  The binary is located 
at a distance $D_{\rm L}=1.25\pm 0.13$ kpc.  The projected separation between the lens 
components is $r_\perp=0.61 \pm 0.07$ AU.  The separation scaled by the mass of the host 
is $r_\perp/M_1=3.2 \ {\rm AU}/M_\odot$.  Under the assumption that separations scale 
with masses, then, the discovered BD is located in the zone of brown-dwarf desert where
BDs are rare.

The lens masses estimated from the $u_0<0$ solution ($M_1=0.15\pm0.02\ M_\odot$ and 
$M_2=0.027\pm 0.003\ M_\odot$) are slightly smaller than those estimated from the $u_0>0$ 
solution due to the smaller value of the measured lens parallax.  We note, however, that 
the difference does not affect the nature of the lens (a BD around a M dwarf).

\section{Summary}

Microlensing provides a powerful probe to study faint or dark objects since these objects 
can be studied independent of their emitted radiation.
By taking advantage of the method, we discovered a brown dwarf 
orbiting a faint star by analyzing the light curve of the gravitational microlensing event 
OGLE-2014-BLG-0257.  Dense resolution of the short-lasting caustic crossing combined with 
the detection of subtle deviation induced by parallax effects, we were able to measure 
both the Einstein radius and the lens parallax.  
This enabled us to unambiguously determine 
the physical parameters of the lens system including the mass, distance, and
projected separation.
It was found that the lens is a binary composed of a brown dwarf and a low-mass star.
It was also found that the brown dwarf companion is located in the zone of the brown dwarf desert.


\acknowledgments
Work by C.~Han was supported by Creative Research Initiative Program (2009-0081561) of 
National Research Foundation of Korea.  
OGLE team thanks Profs.\ M.\ Kubiak and G.\ Pietrzy{\'n}ski, former
members of the OGLE team, for their contribution to the collection of
the OGLE photometric data over the past years.
The OGLE project has received funding from the National Science Centre,
Poland, grant MAESTRO 2014/14/A/ST9/00121 to AU.
Work by AG was supported by JPL grant 1500811.
%
%

\appendix
\section{Xallarap Effects}

\begin{figure}[bht]
\epsscale{1.10}
\plotone{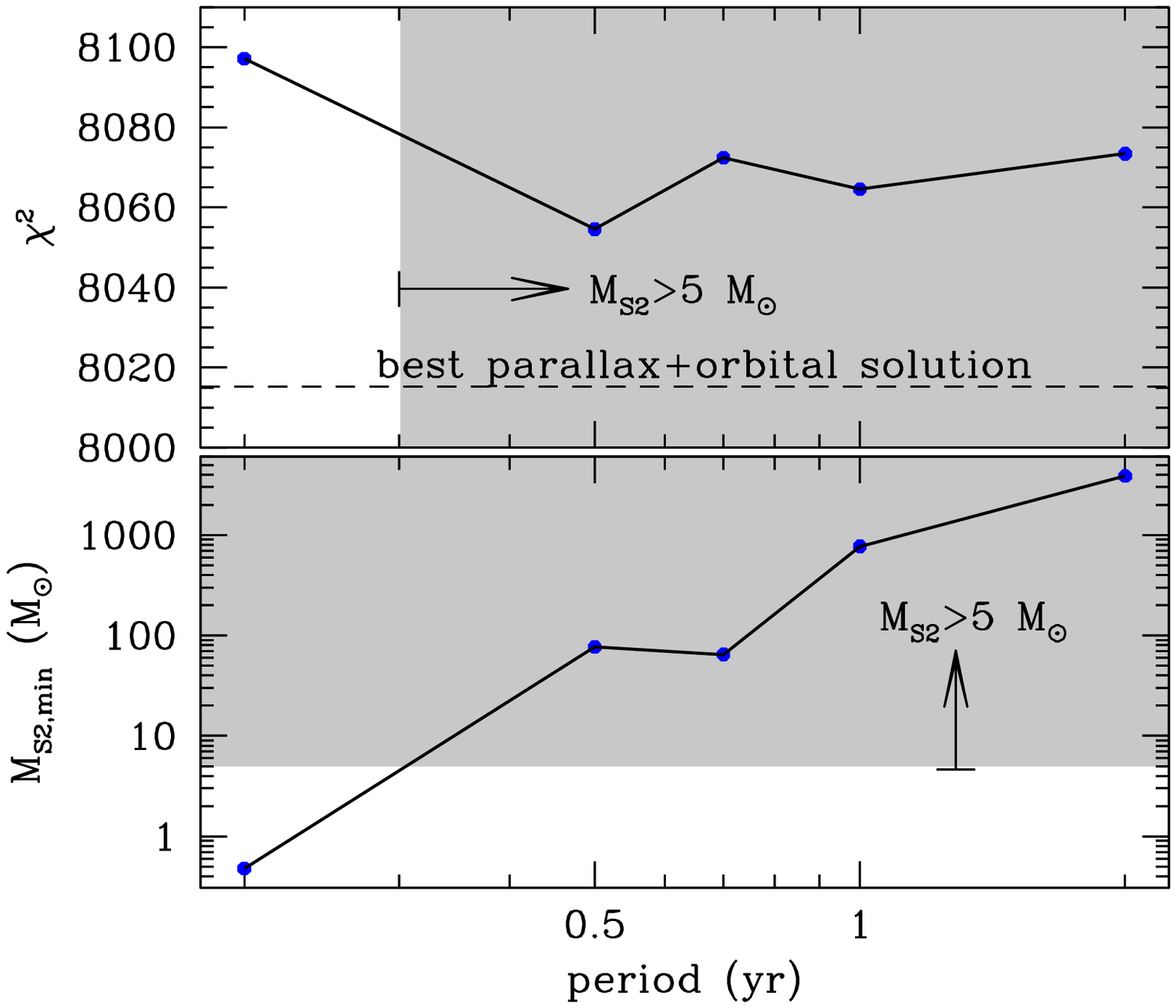}
\caption{\label{fig:a1}
Upper panel: $\chi^2$ distribution of the xallarap fit with respect to the orbital period. 
Lower panel: distribution of the lower mass limit of the source companion.
The dashed line in the upper panel represents the $\chi^2$ value of the 
best-fit orbital + parallax model. 
The shaded areas represent the regions where the estimated minimum mass of the source companion
is greater than $5\ M_\odot$.
}
\end{figure}

It is known that 
long-term deviations in lensing light curves can also be produced by the orbital motion of 
the source star because it is possible for the source orbital motion to mimic parallax 
effects \citep{smith2003, poindexter2005}.  This effect is often referred to as ``xallarap 
effect'', where xallarap is parallax spelled in reverse.  We investigate whether the 
observed long-term deviation can be explained by xallarap effects rather than parallax effects.

In order to consider xallarap effects, one need 5 additional parameters.  These parameters 
include the orbital period $P$, the phase angle and inclination of the orbit, and the north 
and east components of the xallarap vector, $\xi_{{\rm E},N}$ and $\xi_{{\rm E},E}$.
See details about the parameters in \citet{dong2009}.  The magnitude of the xallarap vector 
$\xi=(\xi_{{\rm E},N}^2+\xi_{{\rm E},E}^2)^{1/2}$ corresponds to the major axis of the orbit 
with respect to the center of mass, $a_{\rm S}$, normalized to the projected Einstein radius 
onto the source plane, $\hat{r}_{\rm E}$, i.e. 
\begin{equation}
\xi_{\rm E}={a_{\rm S}\over \hat{r}_{\rm E}}={a_{\rm S}\over D_{\rm S}\theta_{\rm E}},
\label{eqa1}
\end{equation}
where $D_{\rm S}$ is the distance to the source.  The value $a_{\rm S}$ is related to the 
semi-major axis of the orbit by $a_{\rm S}=a M_{{\rm S}2}/(M_{{\rm S}1}+M_{{\rm S}2})$, 
where $M_{{\rm S}1}$ and $M_{{\rm S}2}$ are the masses of the binary source components.
With equation~(\ref{eqa1}) combined with the Kepler's third law, the mass of the source 
companion is expressed in terms of $\xi_{\rm E}$ as 
\begin{equation}
M_{{\rm S}2}={(\xi \hat{r}_{\rm E})^3\over P^2}\left( {M_{{\rm S}1}+M_{{\rm S}2}
\over M_{{\rm S}2}}\right)^2,
\label{eqa2}
\end{equation}
where mass, period, and distance are expressed in $M_\odot$, yr, and AU, respectively.
From the fact that $(M_{{\rm S}1}+M_{{\rm S}2})/M_{{\rm S}2} > 1$, then the the lower mass 
limit of the source companion is expressed as
\begin{equation}
M_{{\rm S}2,{\rm min}}={(\xi \hat{r}_{\rm E})^3  \over P^2}.
\label{eqa3}
\end{equation}

In the upper panel of Figure~\ref{fig:a1}, we present the distribution of $\chi^2$
with respect to the orbital period. In the lower panel, we also present the distribution
of the lower mass limit of the source companion estimated by the relation (\ref{eqa3}).
From the $\chi^2$ distribution, it is found that the best xallarap solution
provides a fit that is worse than the best orbital + parallax solution by $\chi^2=49.4$,
which is statistically important.  Furthermore, the estimated minimum mass of the source 
companion for the best-fit xallarap solution is $M_{{\rm S}2}\sim 100\ M_\odot$, implying 
that the solution is physically unlikely.  Therefore, we conclude that xallarap effects 
cannot explain the observed parallax signal.

\end{document}